\documentclass[12pt]{article}
\usepackage{amssymb}
\usepackage{epsfig}
\usepackage{float}
\usepackage{graphicx}
\usepackage{amsmath}
\newcommand{\nua}[1]{\ensuremath{\rlap
           {\kern-2.5pt\ensuremath
           {\overset{\scriptscriptstyle(-)}{\phantom{\nu}}}}
           {\ensuremath{{\nu}_{#1}}}}}

\begin{document}
\vspace*{4cm}
\begin{center}
PREHISTORY OF NEUTRINO OSCILLATIONS
\end{center}
\begin{center}
S. M. BILENKY
\end{center}
\begin{center}
{\em  Joint Institute for Nuclear Research, Joliot-Curie 6,\\  R-141980, Dubna, Moscow region, Russia}
\end{center}
\begin{abstract}
First ideas of neutrino masses, mixing and oscillations proposed by Bruno Pontecorvo in 1957 and later development of these ideas are considered in some details. Original ideas of the two-neutrino mixing proposed by Maki, Nakagawa and Sakata in 1962 are also discussed.
\end{abstract}

\section{Introduction}
First ideas of neutrino oscillations was proposed by Bruno Pontecorvo in 1957-58
\cite{Pontecorvo:1957cp,Pontecorvo:1957qd}. It was a great time in the particle physics.
\begin{enumerate}
  \item Large violation of invariance under the space inversion $P$ and charge conjugation $C$ was discovered in the $\beta$-decay \cite{Wu:1957my} and $\mu$-decay \cite{Garwin:1957hc,Friedman:1957mz}.
  \item Two-component theory of massless neutrino was proposed by Landau \cite{Landau:1957tp}, Lee and Yang \cite{Lee:1957qr} and  Salam \cite{Salam:1957st}.
  \item The two-component neutrino theory  was impressively confirmed  in the experiment on the measurement of the neutrino helicity \cite{Goldhaber:1958nb}.

\item Feynman and Gell-Mann \cite{Feynman:1958ty}, Marshak and Sudarshan \cite{Sudarshan:1958vf} proposed the universal, $V-A$, current$\times$current weak interaction theory which was in a perfect agreement with all  existed data.
\end{enumerate}
 There was a general belief at that time (and many years later) that {\em neutrino was massless particle.}\footnote{Apparently, this belief was based on the success of the two-component theory and on tritium data in which relatively law (about 100 eV) upper bound of the neutrino mass was obtained.} Let us stress that in the fifties only one type of neutrino was known. This neutrino was discovered in the Reines and Cowan experiment \cite{Cowan:1992xc} via observation of the process $\bar\nu+p\to e^{+}+n$. Today we call it the electron neutrino $\nu_{e}$.
  
According to the two-component neutrino theory existed only left-handed neutrino $\nu_{L}$ and right-handed antineutrino $\bar \nu_{R}$. Transitions $\nu_{L}\rightleftarrows \bar \nu_{R}$
were obviously forbidden.

B. Pontecorvo was impressed  by the idea of
$K^{0}\rightleftarrows\bar K^{0}$ oscillations proposed by
Gell-Mann and Pais in 1955 \cite{GellMann:1955jx}. The basics of
$K^{0}\rightleftarrows\bar K^{0}$ oscillations was the following:
\begin{enumerate}
  \item  $K^{0}$ and $\bar K^{0}$ are particles with different strangeness ($\pm$). These particles  are produced (and detected) in strong interaction processes in which the strangeness is conserved.
  \item Weak interaction does not  conserve the strangeness. Eigenstates of the total Hamiltonian (states with definite masses and widths)  are coherent superpositions\footnote{In the fifties it was assumed that $C$ (and later $CP$) is conserved.
$ |K_{1,2}^{0}\rangle $ are eigenstates of $CP$. Later it was discovered that $CP$ is violated in neutral kaon decays. Nowadays states with definite masses and widths are denoted by
$ |K_{S,L}^{0}\rangle $. They are given by $ |K_{S,L}^{0}\rangle=|K_{1,2}^{0}\pm \epsilon|K_{2,1}^{0}\rangle$ where $|\epsilon|\sim 2\cdot 10^{-2}$. In our discussion we will neglect small effects of the $CP$ violation.}
\begin{equation}\label{K1,2}
|K_{1}^{0}\rangle=\frac{1}{\sqrt{2}}(|K^{0}\rangle + |\bar K^{0}\rangle),\quad    |K_{2}^{0}\rangle=\frac{1}{\sqrt{2}}(|K^{0}\rangle - |\bar K^{0}\rangle).
\end{equation}
\item It follows from (\ref{K1,2}) that $| K^{0}\rangle$ and $|\bar K^{0}\rangle$ states  are "mixed" states:
\begin{equation}\label{mixstat}
| K^{0}\rangle=\frac{1}{\sqrt{2}}(|K_{1}^{0}\rangle+ |K_{2}^{0}\rangle),\quad
|\bar K^{0}\rangle=\frac{1}{\sqrt{2}}(|K_{1}^{0}\rangle- |K_{2}^{0}\rangle)
\end{equation}

\end{enumerate}
In the paper \cite{Pontecorvo:1957cp} B. Pontecorvo  put the following question: "...wheather there exist other "mixed" neutral particles (not necessarily elementary ones) which are not identical to corresponding antiparticles and for which particle $\to$ antiparticle transitions are not strictly forbidden".
He came to the conclusion  that such "mixed" systems could be   muonium $(\mu^{+}-e^{-})$ and antimuonium $(\mu^{-}-e^{+})$ . At that time it was not known that (at least) two different neutrinos ($\nu_{e}$ and $\nu_{\mu}$) exist in nature. In the framework of one neutrino hypothesis transitions $(\mu^{+}-e^{-})\rightleftarrows(\mu^{-}-e^{+})$ are second order in $G_{F}$ allowed transitions ("are induced by the same interaction which is responsible for $\mu$-decay"):
$$ (\mu^{+}-e^{-})\to \nu+\bar\nu\to (\mu^{-}-e^{+})$$
In 1957 paper \cite{Pontecorvo:1957cp} Pontecorvo considered  $(\mu^{+}-e^{-})\rightleftarrows(\mu^{-}-e^{+})$ oscillations in some details. He made in this paper the following remark about neutrino "If the theory of two-component neutrino was not valid (which is hardly probable at present)
and if the conservation law for neutrino charge took no place, neutrino $\to $ antineutrino transitions in vacuum would be in principle possible."

In spite of the problem connected with two-component neutrino theory, in 1957 B. Pontecorvo published the first paper dedicated to neutrino oscillations \cite{Pontecorvo:1957qd}. At that time R.Davis was doing an experiment on the search for  lepton number violating process
$$\bar\nu +^{37}\rm{Cl} \to e^{-}+^{37}\rm{Ar}$$
with reactor antineutrinos. A rumor reached  B.Pontecorvo that Davis observed
such "events". He suggested  that these "events" could be due to transitions of the reactor antineutrinos $\bar\nu_{R}$ into neutrinos $\nu_{R}$ on the way from  reactor to the detector.
In the paper \cite{Pontecorvo:1957qd} B. Pontecorvo wrote "Recently the question was discussed  whether there
exist other
{\em mixed} neutral particles beside the $K^0$ mesons, i.e., particles
that differ from the corresponding antiparticles, with the transitions
between particle and antiparticle states not being strictly forbidden.
It was noted that the neutrino might be such a mixed particle, and
consequently there exists the possibility of real neutrino
$\leftrightarrows$ antineutrino transitions in vacuum, provided that
lepton (neutrino) charge is not conserved.
This means that the neutrino and antineutrino are {\em mixed} particles, i.e., a symmetric and antisymmetric combination of two truly neutral Majorana particles $\nu_1$ and $\nu_2$ of different combined parity".

In other words by analogy with ($K^{0}-\bar K^{0}$) Pontecorvo assumed that
\begin{equation}\label{BPoscil}
|\nu_{R}\rangle
=\frac{1}{\sqrt{2}}(|\nu_{1R}\rangle+ |\nu_{2R}\rangle)\quad
|\bar\nu_{R}\rangle
=\frac{1}{\sqrt{2}}(|\nu_{1R}\rangle- |\nu_{2R}\rangle),
\end{equation}
where $\nu_{1,2}$ are Majorana neutrinos with masses $m_{1,2}$. The mixing  (\ref{BPoscil}) induce  $\bar\nu_{R}\rightleftarrows \nu_{R}$ oscillations.
In order to explain Davis "events"  B.Pontecorvo had to assume that "a definite fraction of particles ($\nu_{R}$) can induce the ($^{37}\rm{Cl}  -^{37}\rm{Ar}$) reaction".

In the paper \cite{Pontecorvo:1957qd} Pontecorvo  pointed out that due to neutrino oscillations in the    Cowan and Reines experiment \cite{Cowan:1992xc}, in which reactor $\bar\nu$'s were detected via the observation of the process $\bar\nu+p\to e^{+}+n$ , {\em a deficit of antineutrino events could be observed.} He wrote
"The cross section of the process $\bar\nu +p\to e^{+}+n$ with
$\bar\nu$ from reactor must be smaller than expected. This is due to the fact that the neutral lepton beam which at the source is capable of inducing the reaction  changes its composition on the way from the reactor to the detector."
And further "It will be extremely interesting to perform C.L. Cowan and F. Reines experiment at different distances from reactor." Pontecorvo concluded  his first paper on neutrino oscillations with the following remark
"Effects of transformation of neutrino into antineutrino and vice versa may be unobservable in the laboratory, but it will certainly occur, at least, on an astronomical scale.''

At the final stage of the Davis experiment  the anomalous candidate events disappeared and only an upper bound for the cross section of the reaction
 $\bar\nu +^{37}\rm{Cl} \to e^{-} +^{37}\rm{Ar}$ was obtained \cite{Davis:1959pba}. B. Pontecorvo soon came to the conclusion that
 $\nu_{R}$ and  $\bar\nu_{L}$, quanta of the right-handed field $\nu_{R}(x)$, could be noninteracting, sterile neutrinos. The terminology "sterile neutrino", which is standard nowadays, was introduced by him in his next paper on neutrino oscillations.

The next paper on neutrino oscillations was written by B. Pontecorvo in 1967 \cite{Pontecorvo:1967fh} after it was known from the Brookhaven experiment \cite{Danby:1962nd} that (at least) two types on neutrinos $\nu_{e}$ and $\nu_{\mu}$ existed in nature. In this paper he considered not only neutrino oscillations but also neutrinoless double $\beta$-decay, the decay $\mu\to e+\gamma$ and other lepton number violating processes.

 In the 1967 paper B. Pontecorvo discussed all possible transitions between
 $\nu_{\mu}$ and $\nu_{e}$. He considered
 $\nu_{eL}\to \bar\nu_{eL}$,  $\nu_{\mu L}\to \bar\nu_{\mu L}$ and other transitions which "transform potentially active particles into particles, which from the point of view of ordinary weak processes are sterile, i.e. practically undetectable". "The only way of observing the effects in question consists in measuring the intensity and time variation of intensity of original active particles". He considered in this paper also oscillations between active  neutrinos: "...there will take place oscillations $\nu_{\mu}\rightleftarrows\nu_{e}$ which in principle are detectable not only by measuring the intensity and time variation of intensity of original particles, but also by observing the appearance of new particles".

In the paper \cite{Pontecorvo:1967fh} B. Pontecorvo  discussed flux of
 solar $\nu_{e}$'s in the case of neutrino oscillations: "From an observational point of view the ideal object is the sun. If the oscillation length is smaller than the radius of the sun region
effectively producing neutrinos, direct
oscillations will be smeared out and unobservable. The only effect on
the earth's surface would be that the flux of observable sun neutrinos
must be two times smaller than the  expected neutrino flux."

Let us stress that this was written {\em before the  results of the Davis solar neutrino experiment \cite{Davis:1968cp} were obtained.} In the Davis experiments it was found that the detected flux of the solar $\nu_{e}$'s was  (2-3) times smaller than the expected flux (the solar neutrino problem). After the paper \cite{Pontecorvo:1967fh} and next Gribov and Pontecorvo  paper \cite{Gribov:1968kq} it was commonly accepted that the  neutrino mixing and oscillations was the most plausible explanation of the solar neutrino problem. Today we know that in order to describe the results of  solar neutrino experiments we need to take into account not only neutrino mixing but also  coherent scattering of neutrino in matter \cite{Wolfenstein:1977ue,Mikheev:1986wj}.

In  the  Gribov and Pontecorvo paper \cite{Gribov:1968kq}  first model of neutrino masses and mixing was developed. Two types of neutrinos  $\nu_{e}$ and $ \nu_{\mu}$ were known at that time. The authors built the scheme without sterile neutrinos:
"...sterile neutrinos should not be considered if it is required that in nature there are only four neutrino states" (left-handed $\nu_{e}$ and $\nu_{\mu}$ and right-handed $\bar\nu_{e}$ and $\bar\nu_{\mu}$). They assumed that
"lepton nonconservation leads to transitions between neutrino states." And further "all possible transitions may be described with the help of an interaction Lagrangian"
\begin{equation}\label{GP}
\mathcal{L}_{I}=-\frac{1}{2} m_{e\bar e}\bar\nu_{eL}(\nu_{eL})^{c}+
m_{\mu\bar \mu}\bar\nu_{\mu L}(\nu_{\mu L})^{c}+m_{e\bar \mu}(\bar\nu_{\mu L}(\nu_{e L})^{c}+\bar\nu_{e L}(\nu_{\mu L})^{c})+\mathrm{h.c.}
\end{equation}
Here $(\nu_{l L})^{c}=C\bar\nu^{T}_{l L}$,~~($C\gamma^{T}_{\alpha}C^{-1}=-\gamma^{T}_{\alpha} $) is  the conjugated field
and $m_{e\bar e},  m_{\mu\bar \mu}, m_{e\bar \mu}$ are real parameters.

After the diagonalization of the Lagrangian (\ref{GP}) they came to the standard mixing relations
\begin{equation}\label{GP1}
\nu_{eL}=\cos\theta \nu_{1L}+\sin\theta \nu_{2L},\quad
\nu_{\mu L}=-\sin\theta \nu_{1L}+\cos\theta \nu_{2L}.
\end{equation}
Here $\nu_{1,2}=C\bar\nu^{T}_{1,2}$  are fields of Majorana neutrinos with masses
\begin{equation}\label{GP2}
m_{1,2}=\frac{1}{2}\left[m_{e\bar e}+m_{\mu\bar \mu}\mp\sqrt{(m_{e\bar e}-m_{\mu\bar\mu})^{2}+4m^{2}_{e\bar \mu}}\right].
\end{equation}
and mixing angle $\theta$ is given by the relation
\begin{equation}\label{GP3}
\tan 2\theta= \frac{2m_{e\bar \mu}}{m_{e\bar e}-m_{\mu\bar \mu}}.
\end{equation}
Gribov and Pontecorvo applied the developed formalism to the solar neutrinos. The cases $m_{e\bar e},m_{\mu\bar \mu}\ll m_{e\bar \mu}$ and
$m_{e\bar e}=m_{\mu\bar \mu}$ they considered as the most attractive. In these cases $\theta=\frac{\pi}{4}$ (maximal mixing) and "neutrino oscillations are similar to the oscillations in the $K^{0}$ beams". If the mixing is maximal "the flux of observable neutrino must be two times smaller than the total sun neutrino flux".

Apparently, analogy with  $K^{0}\rightleftarrows\bar K^{0}$ oscillations  was important for the authors. Strong interaction conserves the strangeness $S$ and weak interaction violates $S$ and induce $K^{0}-\bar K^{0}$ mixing. Analogously, weak interaction conserves  $L_{e}$ and  $L_{\mu}$ and neutrino mixing is induced by some "superweak" interaction (\ref{GP}) which does not conserve  lepton numbers.

We would like to make a remark on the connection of the Gribov-Pontecorvo scheme with a modern status of the neutrino masses and mixing. From the modern point of view the Gribov-Pontecorvo scheme (generalized in \cite{Bilenky:1987ty} ) is based on the lepton number violating Majorana mass term
\begin{equation}\label{Mj}
\mathcal{L}^{M}=-\frac{1}{2}\sum_{l',l} \bar\nu_{l'L}M_{l'l} (\nu_{lL})^{c}+\mathrm{h.c.}=-\frac{1}{2}\sum^{3}_{i=1}m_{i}\bar\nu_{i}\nu_{i}.
\end{equation}
Here $M$ is a symmetrical complex $3\times3$ matrix and $\nu_{i}=\nu^{c}_{i}$ is the field of the neutrino Majorana with the mass $m_{i}$. Let us stress that the {\em Majorana mass term is the most economical mass term}: the left-handed flavor neutrino fields $\nu_{lL}$ which enter into CC and NC enter also into the mass term (there are no other neutrino fields in the Lagrangian). In the case of the Majorana mass term
\begin{itemize}
  \item Neutrinos with definite masses $\nu_{i}$ are Majorana particles.
  \item $\nu_{lL}=\sum^{3}_{i=1}U_{li}\nu_{iL}$ ($l=e,\mu,\tau$). The number of the flavor and massive neutrinos are equal (three). There are no sterile neutrinos.
  \item Neutrino masses $m_{i}$ are parameters. There are no theoretical reasons for their smallness.
\end{itemize}
The most natural and plausible modern approach to the neutrino masses and mixing is based on the dimension five, lepton number violating, non-renormalizable Weinberg effective Lagrangian \cite{Weinberg:1979sa}
\begin{equation}\label{Wein}
\mathcal{L}^{W}=-\frac{1}{\Lambda}\sum_{l',l} (\bar\psi_{l'L}\tilde{\phi})X_{l'l}(\tilde{\phi}^{\dag}\psi_{lL}) +\mathrm{h.c.}.
\end{equation}
Here $\psi_{lL}$ and $\tilde{\phi}$ are the Standard Model lepton doublet and conjugated Higgs doublet, $\Lambda$ is a dimension $M$ constant, which characterizes the scale of a new, beyond the SM physics, and $X$ is a dimensionless matrix. Let us stress that (\ref{Wein}) is the only possible effective Lagrangian which generates the neutrino mass term.

After spontaneous symmetry breaking from the Lagrangian (\ref{Wein}) we come to the Majorana mass term, proposed in 1969 by Gribov and Pontecorvo. The most important difference between the Gribov-Pontecorvo phenomenological approach and the Weinberg effective Lagrangian approach is that neutrino masses $m_{i}$, generated by the effective Lagrangian (\ref{Wein}), are given by the expression
\begin{equation}\label{Wein1}
m_{i}=\frac{v^{2}}{\Lambda}~x_{i}.
\end{equation}
Here $v=(\sqrt{G_{F}})^{-1/2}\simeq 246$ GeV is the electroweak vev and $x_{i}$ is an eigenvalue of the matrix $X$. The Majorana neutrino mass (\ref{Wein1})
has the form of the product of the ratio $\frac{v}{\Lambda}$ and the factor $v~x_{i}$ which has the form of the typical SM mass. It is natural to assume that $\Lambda\gg v$. Thus, the effective Lagrangian mechanism of neutrino mass generation can explain the smallness of neutrino masses with respect to the SM masses of leptons and quarks. The search for  the  neutrinoless double $\beta$-decay of nuclei and  for sterile neutrinos would be crucial tests of this mechanism.\footnote{In a recent interview to "CERN Courier" (November 2017) Weinberg said "...non-renormalizable interaction that produces the neutrino masses is probably also accompanied with non-renormalizable interactions that produce proton decay...We don't  know anything about the details of those terms, but I'll swear they are there"}

I would like now briefly comment  the development of the Pontecorvo's idea of neutrino mixing and oscillations in Dubna. I started a long-term collaboration with Bruno Pontecorvo in 1975. The title of our first paper \cite{Bilenky:1975tb} was "Quark-lepton analogy and neutrino oscillations". At that time it was established that Charged Current of leptons and quarks had the form
\begin{equation}\label{CC}
j_{\alpha}^{CC}=2(\bar\nu_{eL}\gamma_{\alpha}e_{L}+\bar\nu_{\mu L}\gamma_{\alpha}\mu_{L}+\bar u_{L}\gamma_{\alpha}d'_{L}+\bar c_{L}\gamma_{\alpha}s'_{L}).
\end{equation}
Here $d'_{L}$ and $s'_{L}$ are Cabibbo-GIM mixed fields of the $d$ and $s$ quarks
\begin{equation}\label{qmix}
d'_{L}=\cos\theta_{C}d_{L}+\sin\theta_{C}s_{L},\quad s'_{L}=-\sin\theta_{C}d_{L}+\cos\theta_{C}s_{L},
\end{equation}
where $\theta_{C}$ is the Cabibbo angle. It was natural to assume  that neutrinos are also mixed
\begin{equation}\label{numix}
\nu_{eL}=\cos\theta \nu_{1L}+\sin\theta \nu_{2L},\quad
\nu_{\mu L}=-\sin\theta \nu_{1L}+\cos\theta \nu_{2L}
\end{equation}
where $\nu_{1,2}$  are fields of Dirac neutrinos (like quark fields)  with masses $m_{1,2}$. We wrote in the paper \cite{Bilenky:1975tb}:
"In this scheme the neutrinos $\nu_{1}$ and $\nu_{2}$ are described in the same way as the other leptons and quarks (which is perhaps  an advantage of this scheme), whereas in the Gribov-Pontecorvo theory the neutrinos (Majorana) play a special role among the fundamental particles".

 However, we did not see any reasons for the mixing angle $\theta$ to be the same as  the Cabibbo angle $\theta_{C}$. Moreover, we wrote "... the maximal mixing ($\theta=\frac{\pi}{4}$) seems to us the most fruitful hypothesis".

 In the next paper \cite{Bilenky:1976yj} we developed the most general  scheme of neutrino masses and mixing based on the left-handed Gribov-Pontecorvo Majorana mass term, Dirac mass term and right-handed Majorana mass term (the Dirac and Majorana mass term). In this case two  flavor neutrino fields $\nu_{eL}$ and $\nu_{\mu L}$, known at that time, are mixture of  left-handed components of four massive Majorana fields. Assuming that all masses are small we considered in some details transitions of flavor neutrinos into flavor and sterile states and applied the scheme to the solar neutrinos.

In 1978  the first review on neutrino oscillations was written by B. Pontecorvo and me \cite{Bilenky:1978nj}. This review attracted attention of many physicists to the problem of neutrino masses, mixing, oscillations.
The list of papers on  neutrino oscillations was very short at that time\footnote{Except  papers referred above there were also papers \cite{Eliezer:1975ja,Fritzsch:1975rz}.}

 In the review \cite{Bilenky:1978nj} we discussed possible experiments on the search for neutrino oscillations. As an example,  on the search for  neutrino  oscillations in  atmospheric neutrino experiments we wrote: "The averaged neutrino momentum in such experiments is 5-10 Gev and the distance from the neutrino source to the detector is $\simeq 10^{4}$ km for neutrinos coming from the Earth opposite site. ... it is possible to test neutrino mixing hypothesis by comparing the measured and expected $\nu_{\mu}$ intensities. The sensitivities  of such experiments is rather high $\Delta m^{2}\simeq 10^{-3}~\mathrm{eV}^{2}$".

In the end of the seventieth it was known from experiments on the measurement of the high-energy part of the $\beta$-spectrum of $^{3}H$ that neutrino mass was
much smaller than the electron mass (the original Pauli suggestion): $$m_{\beta}\lesssim 10^{-4}~m_{e}$$ Our main question was: do neutrinos have small, nonzero masses? And our main reference theory was the theory of massless, two-component neutrinos. This theory was perfectly confirmed by Goldhaber et al experiment \cite{Goldhaber:1958nb} but, of course, small neutrino masses were not excluded by this experiment.

We had different general arguments in favor of neutrino masses:
\begin{itemize}
  \item there was no principle, like gauge invariance in the case of the photon, which requires that neutrino masses had to be equal to zero,
  \item after the $V-A$ theory, which was based on the assumption that into CC enter left-handed components of {\em all fields}  , it was  natural to assume that neutrinos, like charged leptons, were particles with masses, \footnote{In the sixties B. Pontecorvo discussed the problem of the neutrino mass with L. Landau.  Landau, one of the author of massless two-component neutrino theory, thought at that small neutrino masses was a natural possibility.}
 \item     etc
 \end{itemize}
However, the most important was the understanding, which was clearly expressed in our review, that due to the interference nature of the neutrino oscillations and a possibility to perform experiments at large values of $\frac{L}{E}$ ($L$ is a source-detector distance and $E$ is a neutrino energy) {\em the investigation of neutrino oscillations is the most sensitive way to search for small neutrino masses} (more exactly small neutrino mass-squared differences $\Delta m^{2}$). A condition to observe neutrino oscillations in vacuum has the form
\begin{equation}\label{condition}
\frac{\Delta m^{2}(\mathrm{eV})^{2}L (\mathrm{m})}{4~E(\mathrm{MeV})}\gtrsim 1
\end{equation}
From this condition it followed that different neutrino oscillation experiments(reactor, accelerator, atmospheric, solar) were sensitive to different $\Delta m^{2}$. We stressed in the review that because true values  of the neutrino mass-squared differences
were unknown  {\em it was necessary to search for neutrino oscillations at all neutrino facilities}. As it is well known this strategy finally brought success: neutrino oscillations were discovered in the Super-Kamiokande atmospheric neutrino experiment \cite{Fukuda:1998mi},  in the SNO solar neutrino experiment \cite{Ahmad:2002jz}  in the KamLAND reactor neutrino experiment \cite{Eguchi:2002dm} and in the solar experiments \cite{Cleveland:1998nv,Hirata:1990xa,Anselmann:1994cf,Abdurashitov:1994bc}. The discovery of neutrino oscillations, driven by the atmospheric mass-squared difference, was perfectly confirmed by the accelerator neutrino experiments \cite{Ahn:2006zza,Adamson:2008zt,Abe:2013fuq,Adamson:2016xxw}.

Notice that after it was established via the observation of neutrino oscillations that neutrino masses are different from zero the origin of small neutrino masses became the major  problem.  Our reference theory today is the Standard Model. There exist very convincing arguments that neutrino masses can not be of the same SM Higgs origin as masses of quarks and leptons. Majorana neutrino masses generated by the beyond the SM Weinberg effective  Lagrangian, which we discussed before, apparently is the most plausible possibility.

Let us return back to the history. When Pontecorvo and me were working on the review on neutrino oscillations  our attention was drown to the 1962 Maki, Nakagawa and  Sakata (MNS) paper \cite{Maki:1962mu}   in which two neutrino mixing was considered.
The approach of these authors was based on  the Nagoya model in which proton, neutron and $\Lambda$ were considered as bound states, correspondingly, of neutrino, electron and muon and a vector boson $B^{+}$, "a new sort of matter".

At the time when the paper \cite{Maki:1962mu} was published,  there was an indication that $\nu_{e}$ and $\nu_{\mu}$ are different particles
(from the limit on the probability of the $\mu\to e+\gamma$ decay) but the Brookhaven experiment \cite{Danby:1962nd} was still not finished.

Maki et al introduced  weak neutrinos $\nu_{e}$ and $\nu_{\mu}$ trough the standard leptonic weak current
\begin{equation}\label{StCC}
j_{\alpha}=\bar\nu_{e}\gamma_{\alpha}(1-\gamma_{5}) e+\bar\nu_{\mu }\gamma_{\alpha}(1-\gamma_{5})\mu.
\end{equation}

They wrote in the paper "...neutrinos from which a corresponding barion (say p) should be constructed are not necessary the weak neutrinos themselves; there may be a possibility that the true neutrinos are different from $\nu_{e}$ and $\nu_{\mu}$ but defined by their linear combination"
\begin{equation}\label{MNSmix}
\nu_{1}=\nu_{e}\cos\delta+\nu_{\mu}\sin\delta,
~~\nu_{2}=-\nu_{e}\sin\delta+\nu_{\mu}\cos\delta
\end{equation}
where "...$\nu_{1}$ and $\nu_{2}$ are regarded as the basic particles".

MNS did not consider neutrino oscillations. They wrote "..weak neutrinos are not stable due to occurrence of virtual transitions $\nu_{e}\leftrightarrows\nu_{\mu}$. Therefore, a chain of reactions $\pi^{+}\to \mu^{+}+\nu_{\mu}$, $\nu_{\mu}+A\to (\mu^{-}~ and/or~ e^{-}) +X$ is useful to check the two-neutrino hypothesis if $|m_{\nu_{1}}-m_{\nu_{2}}|<10^{-6}$ MeV  under the conventional geometry of the experiment" (they had in mind the  two-neutrino Brookhaven experiment \cite{Danby:1962nd}). Further they wrote; "Conversely, the absence of $e^{-}$ will be able not only to verify two-neutrino hypothesis but also to provide an upper limit of the mass of the second neutrino $\nu_{2}$".

The modern three-neutrino mixing has the form
\begin{equation}\label{3numix}
    \nu_{lL}(x)=\sum^{3}_{i=1}U_{li}\nu_{iL}(x).
\end{equation}
  In honor of pioneers of ideas of neutrino mixing and neutrino oscillations the $3\times3$ unitary mixing matrix $U$, which is characterized by three mixing angles and one $CP$ phase, is called Pontecorvo-Maki-Nakagawa-Sakata (PMNS) mixing matrix.

Let us return back to neutrino oscillations in vacuum.  Starting from the very first papers, Pontecorvo and his collaborators considered state of flavor neutrino $\nu_{l}$, produced together with charged lepton $l^{+}$  in CC decays , as a superposition of states of neutrinos with definite masses  $\nu_{i}$ with the same momentum and different energies (non stationary state):
\begin{equation}\label{flavstate}
|\nu_{l}\rangle=\sum^{3}_{i=1} U^{*}_{li}~|\nu_{i}\rangle,\quad (l=e,\mu,\tau).    \end{equation}
Here $|\nu_{i}\rangle$ is the state of neutrino with mass $m_{i}$, momentum $\vec{p}$ and energy $E_{i}\simeq p+\frac{m^{2}_{i}}{2p}$. Coherence of the flavor states is ensured by the Heisenberg uncertainty relation and is based on the smallness of neutrino mass-squared differences. Applying to the flavor state the standard evolution operator $e^{-iHt}$ and assuming that initial state is $|\nu_{l}\rangle$ for the neutrino state at the time $t$ we have
\begin{equation}\label{flavstate1}
|\nu_{l}\rangle_{t}=\sum^{3}_{i=1} U^{*}_{li}e^{-iE_{i}t}~|\nu_{i}\rangle=
\sum_{l'=e,\mu,\tau}|\nu_{l'}\rangle(\sum^{3}_{i=1}U_{l'i} e^{-iE_{i}t}U^{*}_{li}).   \end{equation}
From (\ref{flavstate1})  for $\nu_{l}\to \nu_{l'}$ transition probability we obtained the expression
\begin{equation}\label{flavstate2}
P(\nu_{l}\to \nu_{l'})=|\sum^{3}_{i=1}U_{l'i} e^{-iE_{i}t}U^{*}_{li}|^{2}.
\end{equation}
From (\ref{flavstate2}) it follows that  the three-neutrino vacuum transition probability is given by the expression
\begin{equation}\label{flavstate3}
P(\nu_{l}\to \nu_{l'}) =\delta_{l'l}-4\sum_{i>k}
\mathrm{Re}U_{l'i}U^{*}_{li}U^{*}_{l'k}U_{lk}\sin^{2}\Delta_{ki}+2
\sum_{i>k}
\mathrm{Im}U_{l'i}U^{*}_{li}U^{*}_{l'k}U_{lk}\sin2\Delta_{ki},
\end{equation}
which is  standard nowadays.\footnote{The quantum mechanical treatment and understanding of neutrino oscillations is still under active debates (see, for example, \cite{Akhmedov}). It is important to notice that in all neutrino experiments neutrinos are  ultra-relativistic. For the ultra-relativistic neutrinos $t\simeq L$ and different assumptions (same momentum of $\nu_{i}$ and different energies, or same energy and different momenta or different energies and different momenta etc) lead to the same expression (\ref{flavstate3}) for the neutrino transition probability.}
 Here
\begin{equation}\label{flavstate4}
    \Delta_{ki}=\frac{\Delta m_{ki}^{2}L}{4E},\quad \Delta m_{ki}^{2}= m_{i}^{2}- m_{k}^{2}.
\end{equation}
All possible neutrino mass terms, neutrino oscillations in vacuum and in matter, different models of neutrino mixing, neutrinoless double $\beta$ -decay, electromagnetic properties of neutrinos and many other problems were discussed in many details in our review with S. Petcov \cite{Bilenky:1987ty}. This review summarized initial period of the development of the PMNS ideas of neutrino masses and mixing and, apparently, played an important role  in the propaganda of ideas of nonzero neutrino masses, mixing and oscillations.

Possibly, many lessons can be extracted from very rich and  interesting  neutrino history.  I could mention a few of them
\begin{itemize}
  \item Analogy is an important guiding principle (Fermi theory of the $\beta$-decay was based on the analogy with the electromagnetic interaction, B. Pontecorvo idea of neutrino oscillations was based on the analogy with $K^{0}-\bar K^{0}$ oscillations, etc)
  \item Courageous general ideas (not always in agreement with a common opinion) have good chances to be correct (B. Pontecorvo's idea of small neutrino masses at the time when, after the success of the  two-component theory, everybody believed that neutrino is a massless particle).

\item The history of neutrino oscillations is an illustration of a complicated and thorny way of science: publication of courageous pioneer ideas could be inspired by wrong preliminary data (1957 Pontecorvo's paper on the neutrino mixing and oscillation)  or courageous pioneer ideas can be based on wrong models (MNS's idea of the two-neutrino mixing).\footnote{It is interesting that one of the first idea of the seesaw mechanism of the neutrino mass generation \cite{Minkowski:1977sc} was inspired by a preliminary publication of the observation of the $\mu\to e+\gamma$-decay.  Later the observed
    "events" was identified as a background.}

\end{itemize}


\begin{thebibliography}{99}

\bibitem{Pontecorvo:1957cp}
  B.~Pontecorvo,
 {\em Sov.\ Phys.\ JETP} {\bf 6},  429 (1957) [{\em Zh.\ Eksp.\ Teor.\ Fiz. } {\bf 33}, 549 (1957)].

\bibitem{Pontecorvo:1957qd}
  B.~Pontecorvo,
 {\em Sov.\ Phys.\ JETP} {\bf 6},  429 (1957) [{\em Zh.\ Eksp.\ Teor.\ Fiz. } {\bf 33}, 549 (1957)].

\bibitem{Wu:1957my}
  C.~S.~Wu, E.~Ambler, R.~W.~Hayward, D.~D.~Hoppes and R.~P.~Hudson,
 {\em Phys.\ Rev.}  {\bf 105}, 1413 (1957).



\bibitem{Garwin:1957hc}
  R.~L.~Garwin, L.~M.~Lederman and M.~Weinrich, {\em Phys.\ Rev.}  {\bf 105}, 1415 (1957).

\bibitem{Friedman:1957mz}
  J.~I.~Friedman and V.~L.~Telegdi, {\em Phys.\ Rev.}  {\bf 106}, 1290  (1957).

 
\bibitem{Landau:1957tp} L.~D.~ Landau, {\em Nucl.  Phys.}  {\bf 3}, 127 (1957).

\bibitem{Lee:1957qr} T. D. Lee, C. N. Yang, {\em Phys. Rev.}  {\bf 105}, 1671 (1957).

\bibitem{Salam:1957st}
A. Salam, {\em Nuovo Cim.}  {\bf 5}, 299 (1957).

\bibitem {Goldhaber:1958nb} M. Goldhaber, L. Grodzins and A. W. Sunyar, {\em Phys. Rev.} {\bf 109},  1015 (1958).


\bibitem {Feynman:1958ty} R. P. Feynman and M. Gell-Mann, {\em Phys.~Rev.} {\bf 109}, 193 (1958).

\bibitem {Sudarshan:1958vf} E. C. G. Sudarshan and R. E. Marshak, {\em Phys.~Rev.}
{\bf 109}, 1860  (1958).








\bibitem{Cowan:1992xc}
  C.~L.~Cowan, F.~Reines, F.~B.~Harrison, H.~W.~Kruse and A.~D.~McGuire, {\em Science} {\bf 124}, 103  (1956).


\bibitem{GellMann:1955jx}
  M.~Gell-Mann and A.~Pais, {\em Phys.\ Rev.}  {\bf 97}, 1387  (1955).


\bibitem{Davis:1959pba}
  R.~Davis, Jr. and D.~S.~Harmer,
{\em Bull.\ Am.\ Phys.\ Soc.} {\bf 4},  217 (1959).


\bibitem{Pontecorvo:1967fh}
  B.~Pontecorvo, {\em Sov.\ Phys.\ JETP} {\bf 26},  984 (1968) [{\em Zh.\ Eksp.\ Teor.\ Fiz.}  {\bf 53},  1717 (1967)].

\bibitem{Danby:1962nd}
  G.~Danby, J.~M.~Gaillard, K.~A.~Goulianos, L.~M.~Lederman, N.~B.~Mistry, M.~Schwartz and J.~Steinberger,  {\em Phys.\ Rev.\ Lett.} {\bf 9},  36 (1962).






\bibitem{Davis:1968cp}
  R.~Davis, Jr., D.~S.~Harmer and K.~C.~Hoffman,
  {\em Phys.\ Rev.\ Lett.}  {\bf 20}, 1205 (1968).


\bibitem{Wolfenstein:1977ue}
  L.~Wolfenstein,
 {\em  Phys.\ Rev.} {\bf  D 17}, 2369 (1978).

\bibitem{Mikheev:1986wj}
  S.~P.~Mikheev and A.~Y.~Smirnov, {\em Nuovo Cim.}  {\bf C 9},  17 (1986).


\bibitem{Gribov:1968kq}
  V.~N.~Gribov and B.~Pontecorvo,
 {\em Phys.\ Lett.}  {\bf 28B},  493 (1969).

\bibitem{Bilenky:1987ty}
  S.~M.~Bilenky and S.~T.~Petcov, {\em Rev.\ Mod.\ Phys.}  {\bf 59},  671 (1987).
   Erratum: [{\em Rev.\ Mod.\ Phys.}  {\bf 61},  169 (1989)]
   Erratum: [{\em Rev.\ Mod.\ Phys.}  {\bf 60},  575 (1988)]

\bibitem{Weinberg:1979sa}
  S.~Weinberg,
  {\em Phys.\ Rev.\ Lett.}  {\bf 43},  1566 (1979).


\bibitem{Bilenky:1975tb}
  S.~M.~Bilenky and B.~Pontecorvo, {\em  Phys.\ Lett.}  {\bf 61B},  248 (1976).


\bibitem{Bilenky:1976yj}
  S.~M.~Bilenky and B.~Pontecorvo,
  {\em Lett.\ Nuovo Cim.}  {\bf 17},  569 (1976).


\bibitem{Bilenky:1978nj}
  S.~M.~Bilenky and B.~Pontecorvo,
{\em  Phys.\ Rept.}  {\bf 41},  225 (1978).



\bibitem{Eliezer:1975ja}
  S.~Eliezer and A.~R.~Swift, {\em Nucl.\ Phys.}  {\bf  105B},  45 (1976).




\bibitem{Fritzsch:1975rz}
  H.~Fritzsch and P.~Minkowski, {\em Phys.\ Lett.}  {\bf 62B},  72 (1976).

\bibitem{Fukuda:1998mi}
  Y.~Fukuda {\it et al.} [Super-Kamiokande Collaboration],
{\em Phys.\ Rev.\ Lett.}  {\bf 81},  1562 (1998), hep-ex/9807003.


\bibitem{Ahmad:2002jz}
  Q.~R.~Ahmad {\it et al.} [SNO Collaboration], {\em Phys.\ Rev.\ Lett.}  {\bf 89}, 011301  (2002) , nucl-ex/0204008.






\bibitem{Eguchi:2002dm}
  K.~Eguchi {\it et al.} [KamLAND Collaboration],
{\em Phys.\ Rev.\ Lett.}  {\bf 90},  021802 (2003), hep-ex/0212021.


\bibitem{Cleveland:1998nv}
  B.~T.~Cleveland, T.~Daily, R.~Davis, Jr., J.~R.~Distel, K.~Lande, C.~K.~Lee, P.~S.~Wildenhain and J.~Ullman, {\em Astrophys.\ J.}  {\bf 496}, 505 (1998).
 
\bibitem{Hirata:1990xa}
K. S. Hirata \textit{et al.} (Kamiokande), {\em Phys. Rev. Lett.} {\bf 65},  1297 (1990).



\bibitem{Anselmann:1994cf}
  P.~Anselmann {\it et al.} [GALLEX Collaboration], {\em Phys.\ Lett.}  {\bf 327B}, 377  (1994) .

\bibitem{Abdurashitov:1994bc}
  D.~N.~Abdurashitov {\it et al.} [SAGE Collaboration], {\em Phys.\ Lett.}  {\bf 328B} (1994) 234.







\bibitem{Ahn:2006zza}
  M.~H.~Ahn {\it et al.} [K2K Collaboration], {\em Phys.\ Rev.}  {\bf 74D},  072003 (2006), hep-ex/0606032.



\bibitem{Adamson:2008zt}
  P.~Adamson {\it et al.} [MINOS Collaboration], {\em Phys.\ Rev.\ Lett.}  {\bf 101},  131802 (2008), arXiv:0806.2237.

\bibitem{Abe:2013fuq}
K. Abe \textit{et al.} (T2K), {\em Phys. Rev. Lett.} {\bf 111}, 211803 (2013),
arXiv:1308.0465.

\bibitem{Adamson:2016xxw}
P. Adamson \textit{et al.} (NOvA),
{\em Phys. Rev.} {\bf D93}, 051104  (2016), arXiv:1601.05037.


\bibitem{Maki:1962mu}
  Z.~Maki, M.~Nakagawa and S.~Sakata, {\em Prog.\ Theor.\ Phys.}  {\bf 28}, 870 (1962).



\bibitem{Akhmedov} E. Akhmedov,	Proceedings of the conference "History of the Neutrino", September 5-7, 2018, Paris, France.






\bibitem{Minkowski:1977sc}
  P.~Minkowski, {\em Phys.\ Lett.}  {\bf 67B},   421 (1977).
















\end{thebibliography}
\end{document}